\documentclass[nonacm,sigconf]{acmart}
\usepackage[american]{babel}
\usepackage[T1]{fontenc}
\usepackage{times}
\usepackage{url}
\usepackage{xspace}
\usepackage{xcolor}
\usepackage{booktabs}
\usepackage{verbatim}
\usepackage{multirow}

\usepackage{bbding} % Checkmark

\usepackage{titlesec}
\titleformat{\section}[block]{\filcenter\Large\bfseries}{}{}{\MakeUppercase}
\titleformat{\subsection}{\Large\it}{}{}{}
\usepackage[symbol]{footmisc}
\usepackage{bookmark}

\RequirePackage{type1cm}% CM-Fonts including Math-Fonts in all sizes.
\RequirePackage{soul}% Cross out text.
\setstcolor{blue}% Cross out color.

\newcommand{\bscom}[3][]{%
   \noindent
   \st{#2}{\color{blue}\fontsize{8}{8}\selectfont\,#3}%
\ifx#1\empty\else{\color{red}\fontsize{8}{8}\selectfont\,[#1]}\fi
   }

\newcommand{\Ni}{(1)~}
\newcommand{\Nii}{(2)~}

\makeatletter
\newcommand*{\rom}[1]{\expandafter\@slowromancap\romannumeral #1@}
\makeatother

\usepackage{graphicx}
\DeclareGraphicsExtensions{.png,.pdf,.ai,.jpg}
\graphicspath{{./}}

\begin{document}

\title{\huge The Impact of Main Content Extraction on Near-Duplicate Detection}

\author{%
Maik Fr\"obe,\texorpdfstring{$^*$}{} Matthias Hagen,\texorpdfstring{$^*$}{} Martin-Luther-Universit\"at Halle-Wittenberg, Germany \texorpdfstring{\hskip5em}{} %
Janek Bevendorff,\texorpdfstring{$^\dagger$}{}  Michael~V\"olske,\texorpdfstring{$^\dagger$}{} Benno Stein,\texorpdfstring{$^\dagger$}{}Bauhaus-Universit\"at Weimar, Germany \texorpdfstring{\hskip20em}{} \texorpdfstring{\\}{ }
Christopher~Schr\"oder,\texorpdfstring{$^\ddagger$}{} Robby~Wagner,\texorpdfstring{$^\ddagger$}{} Lukas Gienapp,\texorpdfstring{$^\ddagger$}{} Martin~Potthast,\texorpdfstring{$^\ddagger$}{} \texorpdfstring{\\}{ }Leipzig University, Germany%
}

\begin{abstract}
Commercial web search engines employ near-duplicate detection to ensure that users see each relevant result only once, albeit the underlying web crawls typically include (near-)duplicates of many web pages. We revisit the risks and potential of near-duplicates with an information retrieval focus, motivating that current efforts toward an open and independent European web search infrastructure should maintain metadata on duplicate and near-duplicate documents in its index.

Near-duplicate detection implemented in an open web search infrastructure should provide a suitable similarity threshold, a difficult choice since identical pages may substantially differ in parts of a page that are irrelevant to searchers (templates, advertisements, etc.). We study this problem by comparing the similarity of pages for five (main) content extraction methods in two studies on the ClueWeb crawls. We find that the full content of pages serves precision-oriented near-duplicate-detection, while main content extraction is more recall-oriented.

\end{abstract}

{
\titleformat{\section}{\Large\it}{}{}{} % change appearance just for "Abstract"
\maketitle
}

\pagestyle{empty}
\footnotetext{Published at the 3nd International Symposium on Open Search Technology\\
\hspace*{.6cm}(OSSYM 2021), 11-13 October 2021, CERN, Geneva, Switzerland}
\footnotetext[1]{$<$first-name$>$.$<$last-name$>$@informatik.uni-halle.de}
\footnotetext[2]{$<$first-name$>$.$<$last-name$>$@uni-weimar.de}
\footnotetext[3]{$<$first-name$>$.$<$last-name$>$@uni-leipzig.de}

\section{Introduction}
\label{sec:introduction}

Typical web crawls contain many pages with identical or very similar content and different URLs~\cite{fetterly:2003}. Search engines retrieving pages from such web crawls may encounter those near-duplicates in multiple stages of their pipeline. During indexing, omitting near-duplicates might reduce the index size. During retrieval, near-duplicates might occur in the search engine result pages, reducing the user experience because users gain nothing from viewing the same result twice or more on the search engine result pages~\cite{bernstein:2004}. Hence, identifying near-duplicates is a mandatory step in web search, with commercial search engines like Google showing only the ``best'' version from a set of near-duplicates for a query.%
\footnote{\url{developers.google.com/search/docs/advanced/guidelines/duplicate-content}} 

\begin{figure}[tb]%
\centering%
{%
    \includegraphics[width=0.485\textwidth]{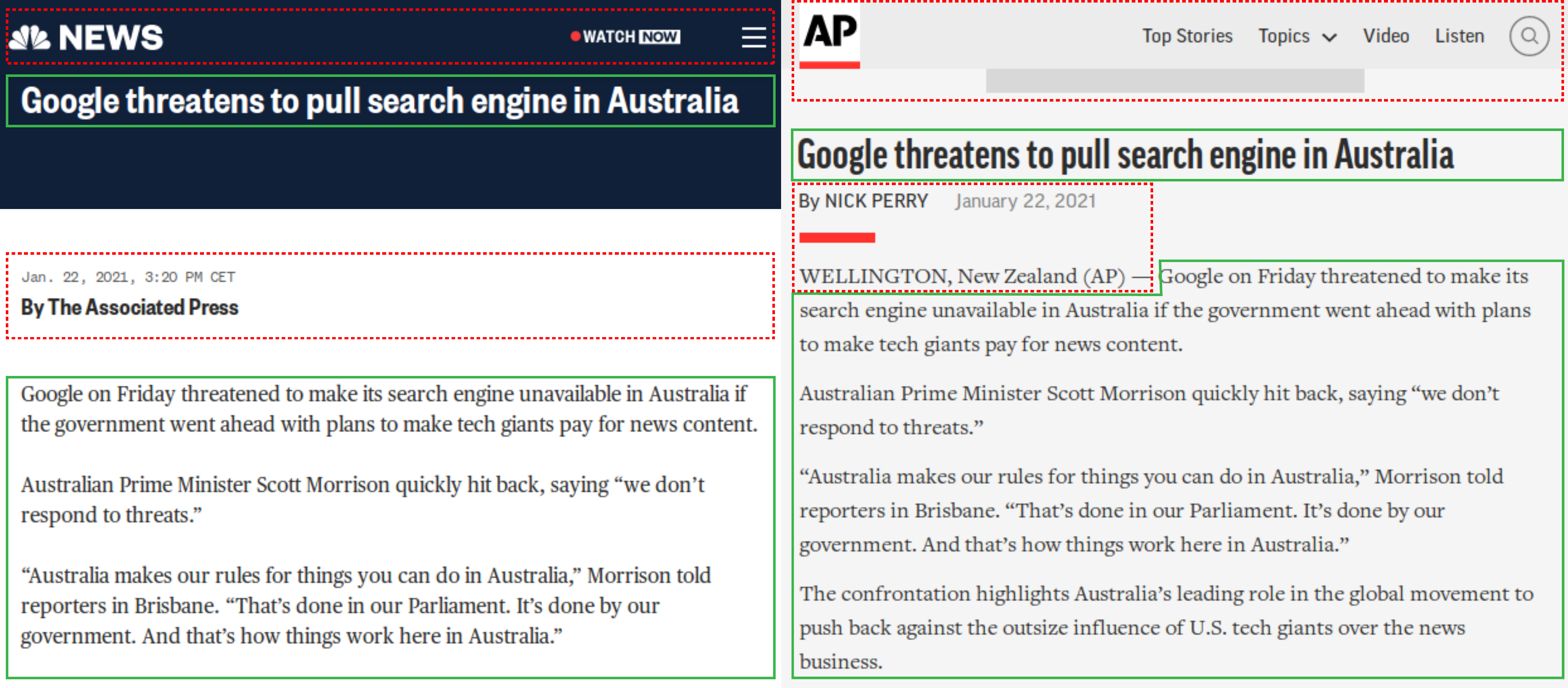}%
}%

\caption{Pages with different URLs and the same article. Both pages have identical content (indicated by green boxes) but vary in parts irrelevant to searchers (indicated by red dashed boxes).}

\label{figure-near-duplicate-example}
\end{figure}

Widely available web crawls---most notably the ClueWebs%
\footnote{\url{lemurproject.org/clueweb09.php/} and \url{lemurproject.org/clueweb12/}} and the Common Crawl%
\footnote{\url{commoncrawl.org/}}---contain the (near-)duplicate documents that the crawler encountered during the crawling process. While the inclusion of near-duplicates enables many applications (like research on text reuse~\cite{alshomary:2019}), it introduces problems for search engines (that we will discuss later in this paper). The CopyCat resource~\cite{froebe:2021} addresses the problems introduced by near-duplicates in information retrieval experiments by providing a precision-oriented near-duplicate detection. CopyCat comes in two parts: \Ni ready-to-use compilations of near-duplicate documents within and between selected web crawls, and \Nii a software library to deduplicate arbitrary document sets, e.g., search engine result pages before they are shown to searchers.

Removing near-duplicates from the search engine result pages with a framework like CopyCat comes with the inherent difficulty of balancing precision and recall. A low precision might reduce the effectiveness of a search engine because relevant and novel documents might be omitted. In contrast, low recall reduces the user experience because users see more near-duplicates. In the context of web search, tuning a similarity measure faces additionally the problem that some parts of the pages that are irrelevant for searchers can increase or decrease the similarity of pages. Figure~\ref{figure-near-duplicate-example} shows an example of two identical articles located at different URLs where the ``noise,'' i.e., the navigation bar, reduces the similarity, eventually having a negative impact on the recall.

Using only the ``retrieval-relevant'' part of documents for the near-duplicate detection might be a promising direction to improve the recall in cases as exemplified in Figure~\ref{figure-near-duplicate-example}. However, the impact of main content extraction on near-duplicate detection is not studied so far. While a ``perfect'' main content extraction should positively affect precision and recall, existing implementations might even harm precision and recall. E.g., invisible changes in the HTML structure of a page might cause tag-based main content extraction approaches to extract different main contents from pages that have identical content from the users' perspective.

We conduct experiments on the ClueWeb crawls to investigate to what extent main content extraction may improve near-duplicate detection in information retrieval. Therefore, we extract document pairs from the ClueWebs containing redundant content (according to their canonical URL) to draw a sharp line between ``roughly similar'' documents and near-duplicates. We calculate the syntactic similarities for all document pairs after extracting the text from the raw HTML with four main content extraction methods and one full content extraction method that does not discard the ``noise'' parts from the document. We evaluate the precision and recall of all five content extraction methods based on manual near-duplicate judgments in two case studies. First, we review document pairs uniformly sampled from the full similarity range for all content extraction methods confirming that main content extraction increases the recall as exemplified in Figure~\ref{figure-near-duplicate-example}. Secondly, we review 100~cases per main content extraction method in which main content extraction changes the similarity drastically, e.g., with identical documents having disjoint main content, or dissimilar pages having duplicate main content, finding inaccurate main content extraction in most of those cases.

\section{Related Work}
\label{sec:related-work}

This section reviews definitions for near-duplicates, their prevalence on the web, and approaches to near-duplicate detection implemented in the CopyCat framework.

\subsection{Defining Near-Duplicates}
The fact that there is no universal near-duplicate definition renders their detection and comparable analysis difficult. Restrictive near-duplicate definitions~\cite{henzinger:2006,manku:2007} consider documents as near-duplicates if they differ only by their session or message IDs, timestamps, visitor counts, server names, invisible differences, URL~parts, or if they are entry pages to the same site. Note that documents, even with minimal content changes, are often not considered near-duplicates under such a restrictive definition. Bernstein and Zobel~\cite{bernstein:2005} relax the near-duplicate definition by applying an information retrieval focus, allowing minimal changes in the content. They consider a document pair as near-duplicate if users get the same information from both documents for all ``reasonable queries.'' We adopt the near-duplicate definition of Bernstein and Zobel since it considers a pair of pages as duplicates if they are equivalent in terms of information provided, i.e., ignoring parts of pages irrelevant to searchers (templates, advertisements,
etc.).

\subsection{Studies on Near-Duplicates on the Web}
According to previous studies by Fetterly et al.~\cite{fetterly:2003,fetterly:2003b}, 30\% of the pages on the web are near-duplicates. While web pages change regularly, consecutive versions of the same web page are usually highly similar~\cite{cho:2000}. Subsequent investigations~\cite{fetterly:2003,fetterly:2003b,ntoulas:2004,adar:2009,olston:2008} confirm this observation by tracking web pages between 5~weeks and one year. For example, Adar et al.~\cite{adar:2009} repeatedly crawl 55,000~URLs over 5~weeks, finding two-thirds of the pages changed their content,  observing that most of these changes were minimal.
Ntoulas et al.~\cite{ntoulas:2004} tracked 150~pages over one year, finding that 40\% of them were still accessible after one year, noticing only insignificant changes on most pages.

\subsection{Near-Duplicate Detection}

There are syntactic, URL-based, and semantic algorithms for detecting near-duplicates~\cite{alsulami:2012}, from which the detection of syntactic near-duplicates received the most attention, resulting in many effective algorithms based on fingerprinting techniques~\cite{broder:1997,charikar:2002,henzinger:2006,manku:2007}.  The CopyCat framework implements syntactic near-duplicate detection in large web crawls with the SimHash algorithm using a fingerprint size of 64~bit and a Hamming-threshold of 3~bits as suggested by Manku et al.~\cite{manku:2007}, while reducing the number of calculated pairwise similarities with the partitioning scheme proposed by Henzinger et al.~\cite{henzinger:2006}. Complementary to estimating the similarity of documents with SimHash, CopyCat can calculate the lossless $S_{3}$ fingerprint similarity~\cite{bernstein:2004} for near-duplicate detection in small sets of documents, such as run and qrel files frequently used in information retrieval experiments.

\section{\hspace*{0.2cm}Near-Duplicates in Web Crawls: Risks and Potentials}
\label{sec:risks-and-potentials}

We recapitulate two risks and one potential of near-duplicate pages in web crawls that the CopyCat framework addresses. Please note that we here focus on information retrieval and that other risks, e.g., in the training of large language models~\cite{raffel:2020}, exist.

\subsection{Risk: Evaluation of Search Engines}
Bernstein and Zobel~\cite{bernstein:2005} found that near-duplicates cause problems in information retrieval evaluations because search engine users do not benefit from seeing near-duplicates. Therefore, they introduce the so-called novelty principle, which states that a document, though relevant in isolation, is irrelevant if it is a near-duplicate to a document the user has already seen on the search engine result page. Especially on web crawls with many near-duplicates, the novelty principle has a non-negligible impact on evaluating search engines~\cite{webis:2020b}. E.g., applying the novelty principle on the runs submitted to the Terabyte track 2004 decreases mean average precision scores by 20\% on average~\cite{bernstein:2005}.

The classical evaluation setup of search engines employs the Cranfield paradigm, making it pretty easy to oversee negative impacts caused by near-duplicates. Relevance assessors judge the relevance of documents to a query in isolation, seeing only one document at a time. Hence, situations that would severely reduce the experience for searchers, e.g., when many near-duplicates occur at subsequent positions in the ranking, can be overlooked because assessors do not look at the ranking. Topic~194 of the ClueWeb09 Web Tracks includes a particularly striking example, where among 47~relevant documents, there are 40 near-duplicates of the same Wikipedia article.

\subsection{Risk: Training of Learning to Rank Models}

Near-duplicates form a kind of oversampling because multiple identical or very similar copies of a page are in the dataset. As recently exemplified~\cite{vandewiele:2020}, oversampling data before partitioning it into training and test sets can invalidate evaluations in machine learning because models may see the same object during training and test. This leakage of information is not possible during the training of learning to rank models because the train/test partitioning is done per query. Still, not removing near-duplicates during the training of learning to rank models decreases the effectiveness of models and biases the trained models~\cite{webis:2020d}.

A study~\cite{webis:2020d} on the ClueWeb09 with 42~ranking features using popular algorithms finds that near-duplicates in the training data harm the retrieval performance, since the presence of near-duplicates is unaccounted for in the loss-minimization of learning to rank and in subsequent evaluations. Furthermore, by varying the number of Wikipedia near-duplicates in the training set, the study showed that models might be biased towards retrieving near-duplicates at higher positions. Hence, these observations make a strong case that learning to rank pipelines benefit from removing duplicate documents from the data before training the model.

Mitigating the negative effects of near-duplicates during the training of retrieval models is easily possible with the CopyCat framework, which can deduplicate the training and test set.

\subsection{Potential: Transfer of Relevance Labels}

In contrast to the previous two risks to the validity and robustness of search engine evaluation and tuning, near-duplicate detection enables the transfer of relevance judgments between different editions (or updates) of web crawls~\cite{froebe:2021}. Relevance judgments---obtained from click logs or expert assessments---are an important and costly resource for the development of search engines. E.g., the effort for the 73,883 relevance judgments for the TREC Web tracks on the ClueWeb09 crawl can be estimated at a manual labor of about 4--8~full-time person-months (assuming 40-hour weeks with 30--60~seconds per judgment~\cite{voorhees:2001}).

To ``reduce'' the costs of keeping the relevance judgments up-to-date for ever-evolving web-indices, search engines might transfer relevance judgments from the previous version of a crawl to the next version when they find the judged documents (or near-duplicates of them) in the newer version of the crawl. In a showcase~\cite{froebe:2021} using precision-oriented near-duplicate detection with the CopyCat framework, 10\% of the ClueWeb09 relevance judgments could be transferred to the ClueWeb12. The number of transferred relevance judgments would even increase to 15\%~when the ClueWeb12 crawling process would have ensured that the URLs judged in the ClueWeb09 are part of the URL seeds for the next crawling round. More frequent updates (compared to the gap of three years in the relevance transfer showcase) would likely further increase the amount of transferrable relevance judgments. Additionally, the reported experiments on the transfer of relevance labels have used only the full content of the pages. Hence, further improvements, e.g., by leveraging main content extraction to increase the recall while maintaining good precision, are possible.

\section{Content Extraction Experiments}
\label{sec:evaluation}

To experimentally compare the impact of main content extraction on near-duplicate detection, we construct a dataset of $186\, 819$~ClueWeb document pairs with redundant content as indicated by canonical URLs. For each document pair, we calculate its syntactic similarity with the lossless $S_{3}$ fingerprinting~\cite{bernstein:2004} for four main content extraction algorithms and the full content of pages. We label 900~document pairs as near-duplicates or not sampled with two approaches: \Ni with 100 document pairs stratified sampled from the $S_{3}$ distribution of each of the five content extraction methods, and \Nii with 50~document pairs with maximal positive/negative $S_{3}$ differences between each of the four main content extraction methods and the full content of a page.

\subsection{Dataset Construction}

We aim at constructing a manageable dataset for our experiments that allows us to draw a sharp line between ``only similar'' documents and near-duplicates. Therefore, we identify document pairs in the ClueWeb09 and ClueWeb12 that should contain redundant content because they share the same canonical URL. We inspect all documents in the ClueWeb, group them by their canonical URL, and select 5,000~groups having the same canonical link at random. From each group, we select all possible pairs (with a maximum of 50 document pairs per group) yielding 186,819~document pairs.

\begin{figure}[tb]%
\centering%
{%
    \includegraphics[width=0.485\textwidth]{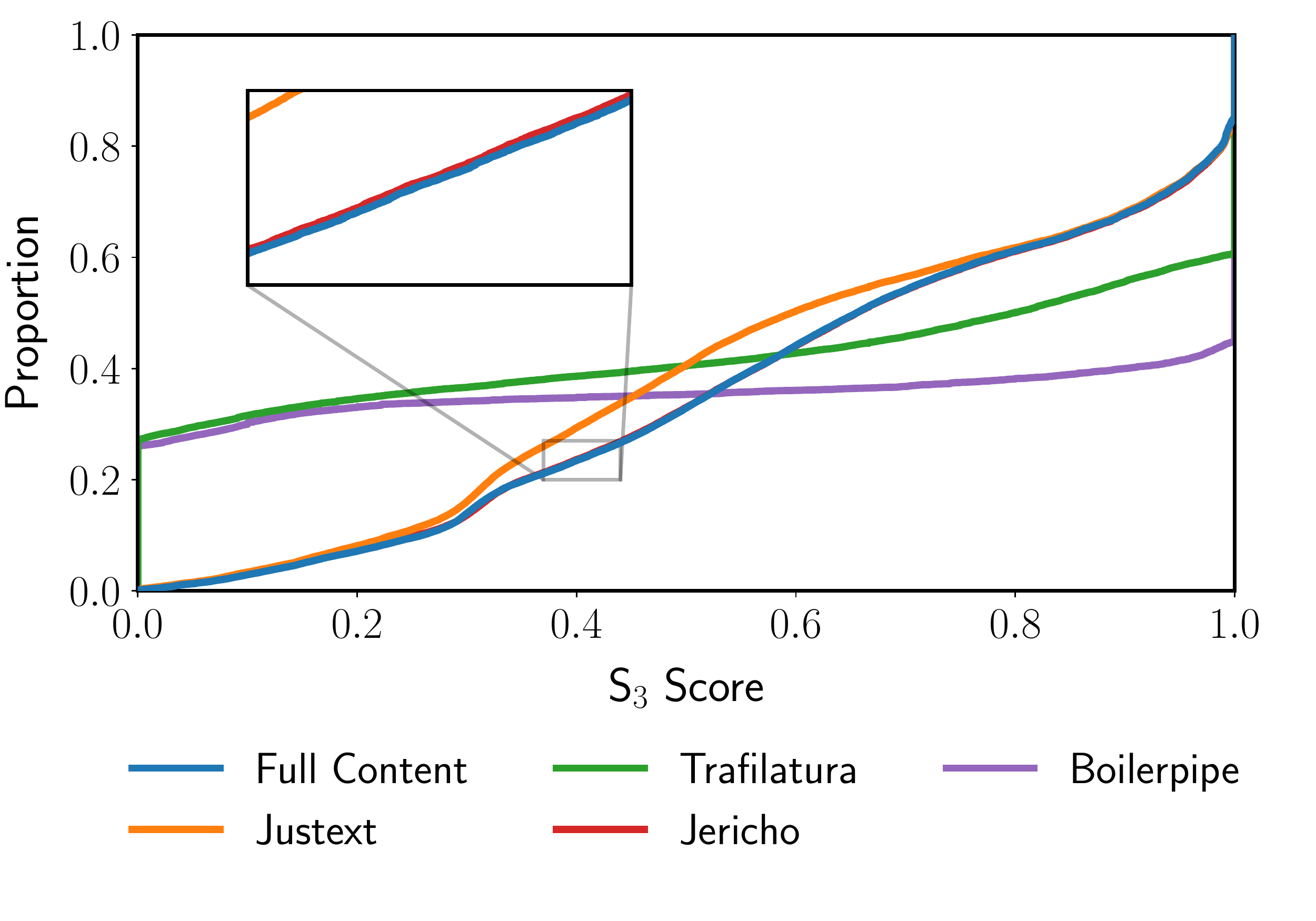}%
}%

\caption{Cumulative distribution plot showing the proportion of document pairs in our dataset below a given similarity measured by their $S_{3}$~score for all five considered content extraction methods.}

\label{figure-s3-score-distribution-ecdf}
\end{figure}

\subsection{Document Preprocessing}

We preprocess all documents with the CopyCat~framework. CopyCat provides five content extraction approaches that transform the raw HTML of a page into text: four main content extraction approaches and one full content extraction. The four provided main content extraction approaches are Boilerpipe~\cite{kohlschuetter:2010}, Jericho,%
\footnote{\url{http://jericho.htmlparser.net}} %
Justext~\cite{pomikalek:2011}, and Trafilatura~\cite{barbaresi:2019}. The full content extraction uses JSoup%
\footnote{\url{https://jsoup.org/}} %
to extract the plain text---without any main content extraction---from the HTML. After extracting the documents text, we remove stop words using Lucene's default stop word list for English, apply stemming with the Porter Stemmer, and lower case the remaining words.

\subsection{Similarity in our Dataset}

We use the lossless $S_{3}$ fingerprint similarity~\cite{bernstein:2004} using word-8-grams to calculate the similarities between all document pairs for all five content extraction methods in our dataset. An $S_{3}$ score of $0$ indicates no overlap between documents, and an $S_{3}$ score of 1 means equality. Figure~\ref{figure-s3-score-distribution-ecdf} shows the cumulative distribution plot for all five content extraction methods regarding the portion of document pairs below a given $S_{3}$ score.

We can identify two groups of content extraction methods that share similar overall behavior. The first group consists of the main content extraction methods Trafilatura and Boilerpipe that show many document pairs with an $S_{3}$~score of~$0$ (26\% for Boilerpipe and 27\% for Trafilatura), indicating that the main content extraction produces disjoint main contents from the considered documents in a pair. This group additionally contains many document pairs with an $S_{3}$ score of~$1$ (55\% for Boilerpipe and 39\% for Trafilatura), indicating that for documents in a pair often the same main content is extracted.

\begin{table*}[tb]
\caption{Overview of (a) precision and recall for near-duplicates in the uniform sampled document pairs at high syntactic similarity ($S_{3} = 1$ and $S_{3} \geq 0.9$), and (b) near-duplicates per $S_{3}$~similarity in the uniform sampled document pairs for all five content extraction methods. Lastly, (c) shows near-duplicates per $S_{3}$~similarity in document pairs with large $S_{3}$~differences to the full content extraction.}
\label{table-precision-recall-content-extraction}

\renewcommand{\arraystretch}{1.25}
\begin{tabular}[t]{@{}l@{\ \ }c@{\ \ }c@{\ \ }c@{\ \ }c@{}}
\multicolumn{5}{@{}l}{(a)} \\[0.75ex]
\toprule
&\multicolumn{2}{@{}c@{\hspace{1em}}}{\bfseries $S_{3} = 1$} & \multicolumn{2}{@{}c@{\hspace{1em}}}{\bfseries $S_{3} \geq 0.9$}\\
\cmidrule(lr){2-3}
\cmidrule(lr){4-5}
& Pr. & Re. & Pr. & Re. \\
\midrule
Full Content & \textbf{1.00} & 0.06 & 0.98 & 0.26 \\
JusText & \textbf{1.00} & 0.07 & \textbf{1.00} & 0.24 \\
Trafilatura & 0.76 & 0.33 & 0.77 & 0.41 \\
Jericho & 0.93 & 0.07 & 0.97 & 0.26 \\
Boilerpipe & 0.60 & \textbf{0.50} & 0.61 & \textbf{0.59} \\
\bottomrule
\end{tabular}
%%%
\hfill
\begin{tabular}[t]{@{}ll@{}}
(b) & (c) \\
\includegraphics[height=123pt,width=171pt]{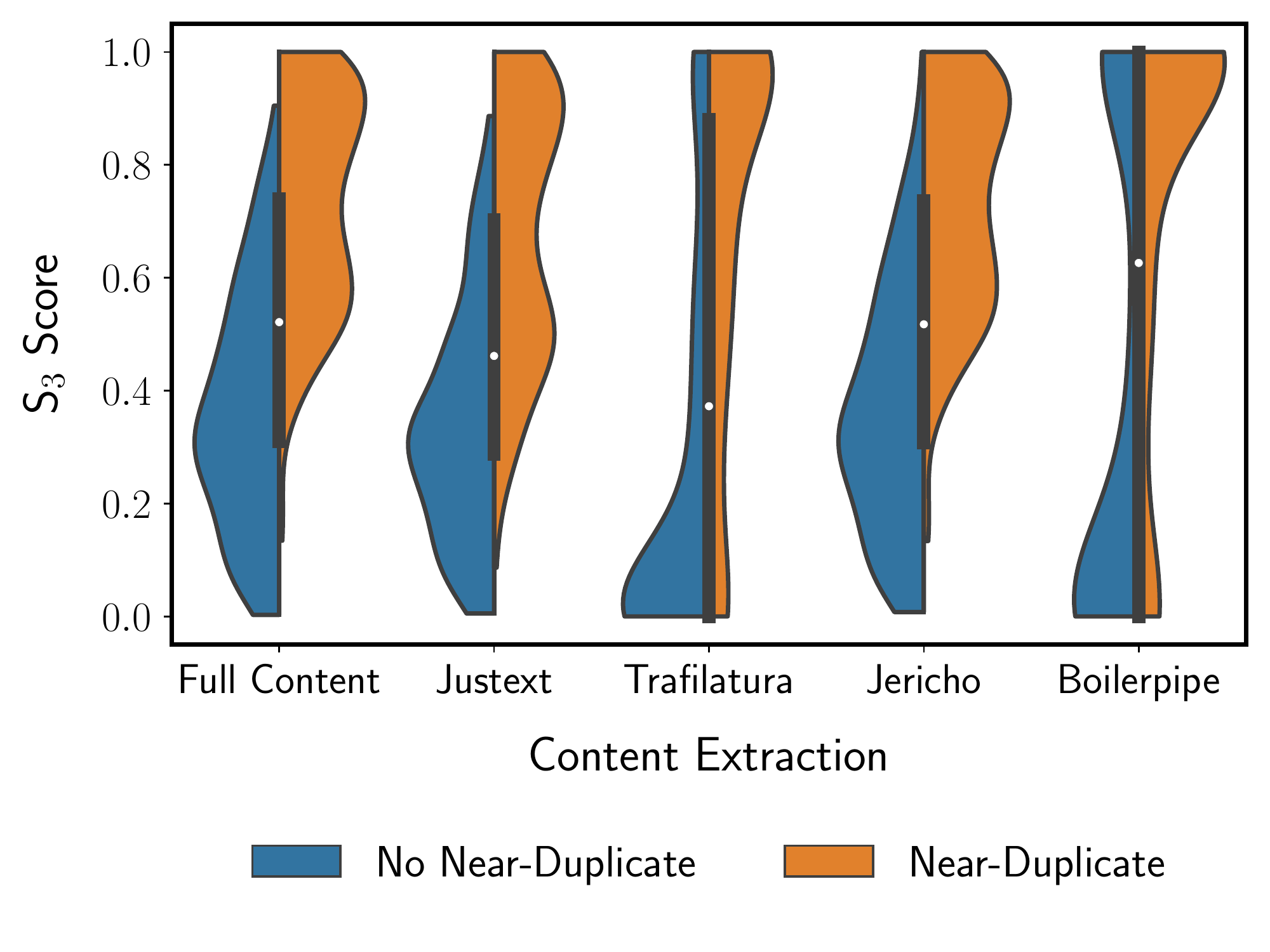} &
\includegraphics[height=123pt,width=171pt]{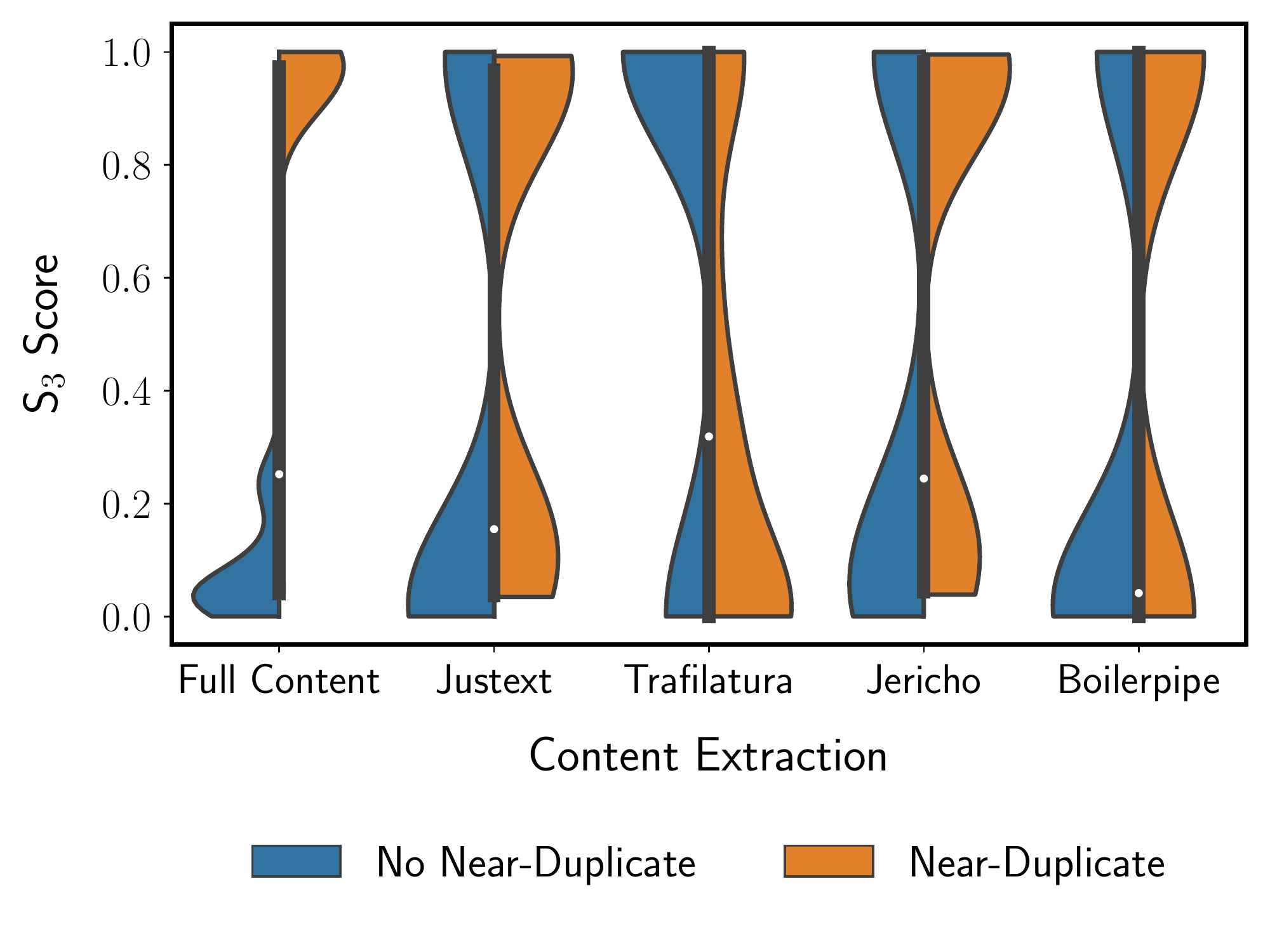} \\[-1ex]
\end{tabular}
 %%%
\hfill
\end{table*}

The second group consists of the full content, Justext, and Jericho methods of content extraction. Approaches in this group have very few document pairs with an $S_{3}$~score of~$0$ (the Justext approach from this group has the maximum of 0.35\% of document pairs with an $S_{3}$ score of~$0$), and much fewer document pairs with an $S_{3}$~score of~$1$ (all three have around~15.6\% of document pairs with an $S_{3}$ score of $1$). We calculated the Pearson correlation between the full content extraction and all other methods. We found a very high correlation to Justext and Jericho (0.97~respectively~0.99) and only a moderate correlation to Boilerpipe and Trafilatura (0.65~respectively~0.73). Overall, Figure~\ref{figure-s3-score-distribution-ecdf} shows that the $S_{3}$ scores in our dataset differ substantially between the two groups, which motivates us to manually verify which of the document pairs are indeed near-duplicates.

\subsection{Labeling Near-Duplicates}

After calculating the $S_{3}$~scores for all document pairs with all content extraction methods, we sample two sets of document pairs for manual review. First, we sample 100~document pairs uniformly covering $S_{3}$ scores between 0~and 1~for all five content extraction methods. Second, we sample document pairs with large $S_{3}$ differences between a main content extraction method and the full content extraction aiming at identifying document pairs where main content extraction yields opposite $S_{3}$ scores to full content extraction. Therefore, we select the 50~document pairs with the largest positive and largest negative $S_{3}$ difference for all four main content extraction methods for manual review.

We use the near-duplicate definition and review guidelines of Bernstein and Zobel~\cite{bernstein:2005} to label near-duplicates: A document pair is considered as near-duplicate when both documents are content-equivalent, and users would be able to extract the same information from either one for all reasonable queries. Two versions of the same Wikipedia article with only minor non-content changes are an example of near-duplicates under this definition.

We labeled the two document pair samples with two assessors. We applied a $\kappa$-test on the 100~document pairs sampled for $S_{3}$ similarities between~$0$ and~$1$ for the full content method, finding a high Fleiss'~$\kappa$ of~0.78, indicating good agreement between both assessors. In a follow-up discussion among the annotators, we discussed all 11~document pairs with different near-duplicate judgments, finally agreeing in all cases. After our $\kappa$~test, each annotator judged the document pairs for the same two main content extraction methods for both user studies.

\subsection{Evaluation}

Table~\ref{table-precision-recall-content-extraction}a and the plot in  Table~\ref{table-precision-recall-content-extraction}b show the ability of all five content extraction methods to identify near-duplicate documents in our set of 500~manually reviewed document pairs that uniformly cover  $S_{3}$ scores between~0 and~1. In Table~\ref{table-precision-recall-content-extraction}a, we report precision and recall for $S_{3}$ thresholds of~1 (for exact duplicates after content extraction) and~0.9 (highly similar extracted content). As in our initial discussion on similarity scores produced by the five content extraction methods, Trafilatura and Boilerpipe (the group with many document pairs at an $S_{3}$ score of 1~in Figure~\ref{figure-s3-score-distribution-ecdf}) as well as the full content, Justext, and Jericho (the group with fewer document pairs with an $S_{3}$ score of 1~in Figure~\ref{figure-s3-score-distribution-ecdf}) show similar behavior in terms of precision and recall. The full content and Justext approaches show a perfect precision of 1.0 at an $S_{3}$ threshold of 1, and Justext even has a perfect precision at an $S_{3}$ threshold of~0.9. On the other side, Trafilatura and Boilerpipe show a very high recall. Even for an $S_{3}$ score of~1, Boilerpipe achieves a remarkable Recall of~0.5.

The plot in Table-\ref{table-precision-recall-content-extraction}b shows the correctly and wrongly identified near-duplicates per $S_{3}$~score for all content extraction methods in our set of 500~manually reviewed document pairs that uniformly cover $S_{3}$ scores between~0 and~1. Again, we can see similar behavior for the full content, Justext, and Jericho methods which make almost no mistakes at high respectively low $S_{3}$ scores. In the opposite group, with Trafilatura and Boilerpipe, we see quite some mistakes (even at $S_{3}$ = 1 and $S_{3}$ = 0).

The plot in Table~\ref{table-precision-recall-content-extraction}c shows the correctly and wrongly identified near-duplicates per $S_{3}$~score for all content extraction methods in our set of 400~manually reviewed document pairs for which the main content extraction changes the similarity drastically. In almost all cases, barring few exceptions, we find that for such large differences, the $S_{3}$ score calculated on the full content correctly identifies near-duplicates and non-near-duplicates. This is visible since the full content method assigns, almost perfectly, non-near-duplicates an $S_{3}$ score near~0, and near-duplicates an $S_{3}$ score near~1. All other approaches make substantial mistakes in this selection of document pairs, indicated by assigning many non-near-duplicates an $S_{3}$ score near~1 (for which the full content method assigned scores near~0, since we selected large differences), and many near-duplicates an $S_{3}$ score near~0 (for which the full content method assigned scores near~1). Especially for cases in which highly similar documents get dissimilar main content extracted, we often found that the main content extraction had problems in identifying the correct main content. Overall, Trafilatura is the most vulnerable in this setting (the most near-duplicates near $S_{3}$ of~0, and most non-near-duplicates near $S_{3}$ of~1). Still, even main content extraction approaches with a very high correlation to the full content extraction method, like Jericho and Justext in our experiments, make substantial mistakes.

\section{Conclusion and Future Work}

We have recapitulated two risks and one potential application of near-duplicates in web search to motivate the maintenance of metadata on duplicate and near-duplicate documents. Given metadata on near-duplicates, it is easy to remove risks such as overestimated evaluation scores of retrieval systems or overfitted learning to rank models. Additionally, updating relevance judgments to the next version of the underlying web crawl can be done at lower costs because relevance labels might automatically be transferred to near-duplicates in the newer version.

In a first attempt to simplify the difficult decision of choosing an appropriate similarity threshold, we investigated how removing parts of documents that are rather irrelevant for the retrieval impacts the similarity of documents. Therefore, we have compared document similarities after preprocessing documents with five (main) content extraction methods. We found that main content extraction can yield very high recall for near-duplicate detection, even when only documents with identical main content are considered as near-duplicates.

An interesting prospect for future work is to include more main content extraction methods and expand the experiments to more document pairs. Another interesting direction for future work might be a further inspection of our observation that highly similar documents having very dissimilar extracted main contents were in most cases caused by mistakes in the main content extraction. This technique might help bootstrap a distant supervision dataset of documents with main content that is difficult to extract.

\begin{raggedright}
\bibliographystyle{abbrv}
\def\bibfont{\fontsize{9pt}{12pt}}
\bibliography{%
  ossym21-near-duplicate-lit%
}
\end{raggedright}

\end{document}